\documentclass[conference]{IEEEtran}%
\usepackage{subfigure,psfrag,graphicx,amssymb,makeidx,multicol,footmisc,subfigure,flushend}
\usepackage[tbtags]{amsmath}
\usepackage{textcomp}
\usepackage[usenames,dvipsnames]{pstricks}
\usepackage{epsfig}
\usepackage{amsfonts}
\usepackage{amssymb}
\usepackage[noadjust]{cite}
\usepackage{graphicx}%
\usepackage{color}
\providecommand{\U}[1]{\protect\rule{.1in}{.1in}}

\DeclareMathOperator*{\tr}{tr}

\include{MyBib.bib}
\begin{document}

\title{Robust Joint Precoder and Equalizer Design in MIMO Communication Systems}
\author{\IEEEauthorblockN{Saeed Kaviani and Witold A. Krzymie\'{n}}
\IEEEauthorblockA{University of Alberta and TRLabs, Edmonton, Alberta, Canada T6G 2V4\\
E-mail: \{saeedk,wak\}@ece.ualberta.ca}}
\maketitle
\begin{abstract}
We address joint design of robust precoder and equalizer in a MIMO communication system using the minimization of weighted sum of mean square errors. In addition to imperfect knowledge of channel state information, we also account for inaccurate awareness of interference plus noise covariance matrix and power shaping matrix. We follow the worst-case model for imperfect knowledge of these matrices. First, we derive the worst-case values of these matrices. Then, we transform the joint precoder and equalizer optimization problem into a convex scalar optimization problem. Further, the solution to this problem will be simplified to a depressed quartic equation, the closed-form expressions for roots of which are known. Finally, we propose an iterative algorithm to obtain the worst-case robust transceivers.
\end{abstract}

\section{Introduction}
Deployment of multiple antennas promises significant capacity gains in wireless systems \cite{Foschini98,Telatar99,Gesbert07,Karakayali06a}, which motivates construction of pragmatic signalling strategies exploiting these gains. Nevertheless, these improvements are severely degraded by interference \cite{Dai04}. Therefore, combination of precoding at the transmitter and equalization at the receiver is often employed to reduce the interference and maximize the performance of system. In this setup, linear strategies have attracted more attention due to their simplicity and robustness. Although various multiple-input multiple-output (MIMO) linear precoding and equalization methods have been proposed \cite{Paulraj01,Scaglione02,Palomar03,Wiesel06}, they assume that the channel state information (CSI) and interference plus noise covariance matrix are perfectly known at the transmitter and receiver. Unavailability of exact information of channel matrices may diminish the performance of the transceivers significantly. This motivates studying of the robust linear transmitter and/or receiver design problem \cite{Vorobyov03,Eldar04,Vucic09,Palomar10c,Wang09,Wang11,Nisar11,Guo05,Guo06,Chiu10,Bogale11,Palomar08}.
\par There are normally two philosophies to consider imperfect CSI and to design robust transmission strategies: Worst-case deterministic scenario and stochastic scenario. In stochastic scenario, the CSI errors are modeled probabilistically and the average performance is optimized \cite{Bogale11,Utschick11a,Palomar08,Joham10}. In this paper, we consider the worst-case scenario since it can characterize the instantaneous imperfect knowledge of system matrices. In this approach, the actual system matrices are assumed to lie within a so-called \textit{uncertainty region} around the estimated values known by the transmitter. A worst-case robust design is a design which achieves a particular performance level for any channel realization staying in the corresponding uncertainty region \cite{Vorobyov03,Eldar04,Guo05,Guo06,Vucic09,Wang09,Palomar10c,Nisar11,Wang11,Chiu10,Zheng08}.

\par Worst-case robust transceiver design has been recently considered in \cite{Wang11,Vucic09}. In \cite{Vucic09}, joint optimization of transceivers is addressed using semi-definite program (SDP) reformulation. However, even for the case of perfect channel knowledge it only gives a suboptimal solution \cite{Kaviani11b}. Moreover, SDP-based approaches (see for example \cite{Vorobyov03,Eldar04}) do not give a closed-form solution and the resultant algorithms require solving a SDP at each iteration. Closed-form solution for the worst-case robust MMSE precoder assuming pre-fixed equalizer is given in \cite{Palomar10c} and it is extended to design of worst-case robust MMSE transceivers in \cite{Wang11}. Nevertheless, the proposed algorithm is based on alternative optimization between precoder and equalizer (rather than joint optimization). It also involves solving a quintic equation, for which a closed-form solution is unknown and it is only solved numerically.
\par Conventionally, the robust transmission strategies in the MIMO communication systems consider imperfect knowledge of channel gains between the transmit and receive antennas. Here, we consider a wider range of system parameters \textit{all} known inaccurately to the system. In addition to CSI, we consider imperfect knowledge of interference plus noise covariance matrix, and power shaping matrix. 
Moreover, our objective is weighted sum of mean square error minimization which is a more general performance metric \cite{Luo11}. We avoid SDP reformulation to solve the problem and consequently the proposed algorithm has lower complexity. First, we obtain the worst-case system matrices within the uncertainty region. Then, we \textit{jointly} optimize precoder and equalizer where the resultant matrix-valued optimization problem is reduced to a scalar convex problem. Further, the solution to this problem can be simplified to a \textit{depressed quartic equation}. Interestingly, the solution of a quartic equation can be expressed in terms of radicals. Hence, a closed-form expression for the precoder and equalizer matrices can be obtained. Finally, we propose an iterative algorithm to find the optimal transceivers. Note that the analysis and design approach for point-to-point MIMO systems presented in this paper can be extended to multiuser MIMO systems \cite{Kaviani11c}.

\par \textit{Notation:} We denote positive semi-definite matrices as $\mathbf A \succeq \mathbf 0$. Capital bold letters represent matrices and small bold letters represent vectors. We denote conjugate transpose (Hermitian) operator with $(\cdot)^\mathsf H$. $\mathbf A^{-\frac{1}{2}}$ represents the inverse square root of positive definite matrix $\mathbf A$. $\tr\left\{\cdot\right\}$ is matrix trace operator and $\| \cdot \|$ demonstrates Frobenius norm of a matrix. $\lambda_{\max}(\cdot)$ is reserved to denote the maximum eigenvalue of a matrix.

\section{System Model and Preliminaries}
We consider a MIMO communication system, where the transmitter is equipped with $n_t$ antennas and the receiver employs $n_r$ antennas ($n_r \leq n_t$). The transmitter broadcasts a data vector denoted by $\mathbf u\in \mathbb C^{n_r}$ using the linear precoding matrix $\mathbf F \in \mathbb C^{n_t\times n_r}$. The channel between the transmitter and the receiver is characterized by the matrix $\mathbf H \in \mathbb C^{n_r \times n_t}$. The receiver observes the signal
\begin{equation}
\mathbf y=\mathbf H\mathbf F\mathbf u+\mathbf n,
\end{equation}
where $\mathbf n$ represents the correlated interference plus noise vector. We denote the interference plus noise covariance matrix as $\mathbf \Omega=\mathbb E\left[\mathbf n\mathbf n^\mathsf H\right] \in \mathbb C^{n_r \times n_r}$. The linear processing at the receiver can be characterized by the equalizer matrix $\mathbf G \in \mathbb C^{n_r \times n_r}$. Hence, the estimated symbol vector at the receiver can be described as
\begin{equation}
\hat{\mathbf{u}}=\mathbf{G}\mathbf{y}.
\label{eqn:Estimate}%
\end{equation}
Let us define the estimation error covariance matrix as
\begin{equation}
\begin{split}
\mathbf E=&\mathbb E\left[\left(\hat{\mathbf u}-\mathbf u\right)\left(\hat{\mathbf u}-\mathbf u\right)^\mathsf H\right]\\=&\mathbf G\mathbf{H}\mathbf F%
\mathbf F^{\mathsf{H}}\mathbf{H}^{\mathsf{H}}\mathbf G^\mathsf H%
-\mathbf G\mathbf{H}\mathbf F-\mathbf F^{\mathsf{H}}\mathbf{H}^{\mathsf{H}}\mathbf G^{\mathsf{H}}
+\mathbf G\mathbf{\Omega}\mathbf G^{\mathsf{H}}%
+\mathbf{I}.%
\label{eqn:MSE_matrix2}
\end{split}
\end{equation}
which is referred to as \textit{mean square error (MSE)-matrix} \cite{Palomar03}. We are specifically interested in the following problem:
\begin{equation}
\begin{array}[c]{rl}
\underset{\mathbf G, \mathbf F}{\text{minimize}} & \tr\left\{\mathbf W\mathbf E\right\}\\
\text{subject to} & \tr\left\{\mathbf \Phi\mathbf F\mathbf F^\mathsf H\right\} \leq P
\end{array}
\label{eqn:optimization3}
\end{equation}
where the optimization is over precoding and equalization matrices with given diagonal weight matrices $\mathbf{W}\in\mathbb{C}^{n_r\times
n_r}$ where the main diagonal of $\mathbf{W}$ is denoted by $[w_{1}%
,...,w_{n_r}]$ with non-negative weights $w_{j}\geq0.$ Since the diagonal elements of the MSE-matrix are MSE values of the estimated symbol vector, the problem (\ref{eqn:optimization3}) is often called \textit{weighted sum of mean square error minimization} \textit{(WMMSE)} problem (see \cite{Christensen08,Luo11,Kaviani11a,Kaviani11b} for details). We also account for a linear power constraint $\tr\left\{\mathbf \Phi\mathbf F\mathbf F^\mathsf H\right\}\leq P$ and specifically refer to the weight matrix $\mathbf \Phi$ in the power constraint as the \textit{power shaping matrix}. This matrix also can characterize the direction, in which the transmitted power can propagate while reducing the interference in other directions (e.g. to other users in a multiuser case). Additionally, we assume that the matrix $\mathbf \Phi$ is full rank and square with size of $n_t$. This assumption is a practical constraint due to the fact that if $\mathbf \Phi$ is a rank deficient matrix then one can always transmit infinite power in one direction without violating the power constraint. Notice that when $\mathbf \Phi=\mathbf I$, the sum power constraint emerges. We refer to the matrices $\mathbf H,\mathbf \Phi,$ and $\mathbf \Omega$ as \textit{system matrices}.

\par It is shown that any performance metric characterized by some particular function $f\left(\mathbf{E}\right)$ of the MSE-matrix $\mathbf{E}$, can be approximated using the problem (\ref{eqn:optimization3}) \cite{Christensen08,Paulraj01,Kaviani11b}. The approach is that at each iteration, we select $\mathbf W=\nabla_{\mathbf E}f(\bar{\mathbf E})^\mathsf T$ at the operating point $\overline{\mathbf E}$, then solve the optimization problem (\ref{eqn:optimization3})\footnote{The optimal forms of precoder and equalizer diagonalize the MSE-matrix $\mathbf E$ \cite{Kaviani11b} and consequently the updated weight matrix $\mathbf W$ is diagonal.}. The algorithm iterates until the convergence is achieved. For example, to adopt sum rate maximization one can select $\mathbf W=\bar{\mathbf E}^{-1}$ at each iteration. This approach has been extensively used in \cite{Christensen08,Paulraj01,Schmidth09,Kaviani11b,Luo11} to optimize any performance function of the MSE-matrices (e.g. sum-rate), since the resultant problem becomes convex by fixing any optimization variable.

\section{Perfect Knowledge of System Matrices}
We begin with the case of perfect channel knowledge which will be employed for the robust design. \cite{Christensen08,Luo11} have considered WMMSE problem with perfect CSI, but their solutions for each of the precoding and equalization matrices are interdependent (requires alternative optimization). Here, we give an extension of the results in \cite{Paulraj01,Palomar03} when power shaping matrix is present. Detailed discussion of this problem with more general constraints is available in \cite{Kaviani11b}. Here, a special case of this result where the number of transmitted data streams is equal to $n_r$ is given.
\par \textit{Lemma 1 \cite{Kaviani11b}:} For any channel matrix $\mathbf H$ and given the full rank and square matrices $\mathbf \Phi$ and $\mathbf \Omega$, the optimum precoding and equalization matrices of the problem (\ref{eqn:optimization3}) have the following structure
\begin{align}
\mathbf F=& \mathbf \Phi^{-\frac{1}{2}}\mathbf V\mathbf \Sigma,\label{eqn:F}
\\
\mathbf G=&\mathbf \Lambda\mathbf U^\mathsf H\mathbf \Omega^{-\frac{1}{2}}.\label{eqn:G}
\end{align}
where $\mathbf \Sigma$ and $\mathbf \Lambda$ are diagonal matrices with the diagonal elements $\sigma_i\geq 0$ and $\lambda_i\geq 0, i=1,\ldots,n_r$, respectively. $\mathbf U \in \mathbb C^{n_r \times n_r}$ and $\mathbf V \in \mathbb C^{n_t \times n_r}$ are obtained by performing the singular value decomposition (SVD) of the following matrix
\begin{equation}
\mathbf \Omega^{-\frac{1}{2}}\mathbf H\mathbf \Phi^{-\frac{1}{2}}=\mathbf U \left[\mathbf \Gamma \quad \mathbf 0_{n_r \times n_t-n_r}\right] \left[\mathbf V \quad \breve{\mathbf V}\right]^\mathsf H, \label{eqn:SVD}
\end{equation}
in which $\mathbf \Gamma$ contains its $n_r$ nonzero eigenvalues $\gamma_1 \geq \ldots \geq \gamma_{n_r}$ and $\breve{\mathbf V} \in \mathbb C^{n_t \times (n_t-n_r)}$ contains the right singular vectors corresponding to the zero eigenvalues\footnote{The matrix $\mathbf \Omega^{-\frac{1}{2}}\mathbf H\mathbf \Phi^{-\frac{1}{2}}$ with probability one has a rank of $n_r$, due to the random nature of the channel matrix $\mathbf H$ and the fact that $n_r \leq n_t$.}.
\begin{IEEEproof}
The proof can be found in \cite{Kaviani11c}.
\end{IEEEproof}

\section{Imperfect Knowledge of System Matrices}\label{Sec:ImperfectCSI}
\par In this case, only estimated matrices $\widehat{\mathbf H}$ and $\widehat{\mathbf \Omega}$ and $\widehat{\mathbf \Phi}$, are available at the transmitter and receiver. Therefore, the actual value of these matrices can be described as a sum of the estimated matrices and the error matrices:
\begin{align}
\mathbf H&=\widehat{\mathbf H}+\mathbf \Delta_{H},\label{eqn:H}
\\
\mathbf \Omega&=\widehat{\mathbf \Omega}+\mathbf \Delta_{\Omega},\label{eqn:Omega}
\\
\mathbf \Phi&=\widehat{\mathbf \Phi}+\mathbf \Delta_{\Phi}\label{eqn:Phi}.
\end{align}
We are interested in the joint optimization of the precoder and equalizer, while the unknown actual system matrices are guaranteed to fit in the (norm-based) uncertainty region. Hence, the error region can be described as
\begin{align}
\mathcal U=\Bigl\{ (\mathbf \Delta_{H},\mathbf \Delta_\Omega,\mathbf \Delta_\Phi): \|\mathbf \Delta_{H}\|\leq \varepsilon_{H}, \|\mathbf \Delta_\Omega\|\leq \varepsilon_\Omega,\Bigr. \nonumber\\
\Bigl.\widehat{\mathbf \Omega}+\mathbf \Delta_\Omega \succeq \mathbf 0, \|\mathbf \Delta_{\Phi}\|\leq \varepsilon_\Phi, \widehat{\mathbf \Phi}+\mathbf \Delta_\Phi \succeq \mathbf 0 \Bigr\}.
\end{align}
Consequently, the worst-case transceiver design can be expressed as
\begin{equation}
\begin{array}
[c]{rl}%
\underset{\mathbf F,\mathbf G}%
{\text{minimize }}&\underset{\left(\mathbf \Delta_{H},\mathbf \Delta_{\Phi},\mathbf \Delta_\Omega\right) \in \mathcal U}{\text{ max}} \tr\left\{\mathbf W\mathbf E\right\}\\
\text{subject to} & \tr\left\{%
\mathbf \Phi\mathbf F\mathbf F^\mathsf{H}\right\}  \leq P.
\end{array}
\label{eqn:optimization5}
\end{equation}
Please note that we consider a case, in which the uncertainty of the system matrices is the same at the transmitter and the receiver. We leave the scenario, under which the uncertainty of the system matrices at the transmitter is much higher than at the receiver to future work.
\subsection{Finding Least Favorable System Matrices}
We proceed by finding the worst-case estimation errors for the system matrices. First, we expand the objective function of (\ref{eqn:optimization5}) in terms of the estimated and error system matrices using the definitions (\ref{eqn:MSE_matrix2}) and (\ref{eqn:H})-(\ref{eqn:Phi}) and simplify the worst-case problem as
\begin{equation}
\begin{array}
[c]{rl}%
\underset{\left(\mathbf \Delta_{H},\mathbf \Delta_{\Phi},\mathbf \Delta_\Omega\right) \in \mathcal U}{\text{maximize}} & \tr\left\{\mathbf W\widehat{\mathbf E}\right\}+\tr\left\{\mathbf G^\mathsf H\mathbf W\mathbf G\mathbf \Delta_\Omega\right\}
\\
& +\tr\left\{\mathbf A\mathbf \Delta_{H}\mathbf B\mathbf \Delta_{H}^\mathsf H\right\}+2\mathfrak{Re}\left\{\tr\left\{\mathbf C\mathbf \Delta_{H}\right\}\right\}\\ 
\text{subject to} & \tr\left\{%
\widehat{\mathbf \Phi}\mathbf F\mathbf F^\mathsf{H}\right\}  \leq P-\tr\left\{\mathbf \Delta_\Phi\mathbf F\mathbf F^\mathsf H\right\}
\end{array}
\label{eqn:optimization6}
\end{equation}
where $\widehat{\mathbf E}$ is the MSE-matrix defined in (\ref{eqn:MSE_matrix2}) based on the estimated matrices $\widehat{\mathbf H}$ and $\widehat{\mathbf \Omega}$ and
\begin{align}
\mathbf A=&\mathbf G^\mathsf H\mathbf W\mathbf G,\\
\mathbf B=&\mathbf F\mathbf F^\mathsf H,\\
\mathbf C=&\mathbf F\mathbf F^\mathsf H \widehat{\mathbf H}^\mathsf H\mathbf G^\mathsf H\mathbf W\mathbf G-\mathbf F\mathbf W \mathbf G.
\end{align}

\par \textbf{Least Favorable Matrices $\mathbf \Delta_\Omega$ and $\mathbf \Delta_\Phi$:}
Since the error matrices are independent of each other, the least favorable interference plus noise covariance matrix can be obtained from the problem
\begin{equation}
\begin{array}[c]{rl}
\underset{\|\mathbf \Delta_\Omega\|\leq \varepsilon_\Omega}{\text{maximize}} & \tr\left\{\mathbf G^\mathsf H\mathbf{WG\Delta}_{\Omega}\right\}\\
\text{subject to} & \widehat{\mathbf \Omega}+\mathbf \Delta_\Omega \succeq \mathbf 0.
\end{array}
\label{eqn:optimization7}
\end{equation}
We assume that $\varepsilon_\Omega$ is small enough to ignore the positive semi-definite condition of the problem (\ref{eqn:optimization7}). Nevertheless, we will see that this relaxation gives us a solution, which also satisfies a positive semi-definite condition.
Using Cauchy-Schwartz inequality, we can obtain
\begin{equation}
\tr\left\{\mathbf G^\mathsf H\mathbf{WG\Delta}_{\Omega}\right\} \leq \|\mathbf G^\mathsf H\mathbf W\mathbf G\|\cdot \|\mathbf \Delta_\Omega\|\leq \varepsilon_\Omega\|\mathbf G^\mathsf H\mathbf W \mathbf G\|
\end{equation}
and the upper bound occurs when
\begin{equation}
\mathbf \Delta_\Omega^\star=\varepsilon_\Omega\frac{\mathbf G^\mathsf H\mathbf W\mathbf G}{\|\mathbf G^\mathsf H\mathbf {WG}\|}.
\end{equation}
which gives the worst-case estimation error matrix $\mathbf \Delta_\Omega^\star$.
\par We continue by finding the worst-case estimation error of the interference direction matrix, i.e. $\mathbf \Delta_\Phi$. It is trivial that the worst-case happens when the maximum allowed power is minimized. Consequently, we are interested in this optimization problem:
\begin{equation}
\begin{array}[c]{rl}
\underset{\|\mathbf \Delta_\Phi\|\leq \varepsilon_\Phi}{\text{maximize}} & \tr\left\{\mathbf \Delta_\Phi\mathbf F\mathbf F^\mathsf H\right\}\\
\text{subject to} & \widehat{\mathbf \Phi}+\mathbf \Delta_\Phi \succeq \mathbf 0
\end{array}
\label{eqn:optimization8}
\end{equation}
Similarly to the problem (\ref{eqn:optimization7}), we can obtain the worst-case error matrix as
\begin{equation}
\mathbf \Delta^\star_\Phi=\varepsilon_\Phi\frac{\mathbf F\mathbf F^\mathsf H}{\|\mathbf F\mathbf F^\mathsf H\|}.
\end{equation}
Substituting these worst-case estimation errors $\mathbf \Delta^\star_\Phi$ and $\mathbf \Delta^\star_\Omega$ into the problem (\ref{eqn:optimization5}) results in the terms $\varepsilon_\Omega\|\mathbf G^\mathsf H\mathbf {WG}\|$ and $\varepsilon_\Phi\|\mathbf {FF}^\mathsf H\|$. 
Since we are interested in the worst-case scenario, we can use upper bounds of these terms alternatively. Hence, we can write inequalities
\begin{align}
\varepsilon_\Omega\|\mathbf G^\mathsf H\mathbf {WG}\|\leq &\varepsilon_\Omega\|\mathbf W^{\frac{1}{2}}\mathbf G\|^2=\varepsilon_\Omega\tr\left\{\mathbf G^\mathsf H\mathbf W\mathbf G\right\} \label{eqn:DeltaOmegaTerm}
\\
\varepsilon_\Phi\|\mathbf F\mathbf F^\mathsf H\|\leq &\varepsilon_\Phi\|\mathbf F\|^2=\varepsilon_\Phi\tr\left\{\mathbf F\mathbf F^\mathsf H\right\}.\label{eqn:DeltaPhiTerm}
\end{align}
where we have used the inequality $\|\mathbf X\mathbf Y\|\leq \|\mathbf X\|\cdot\|\mathbf Y\|$ which can be proved by utilizing the Cauchy-Schwartz inequality \cite{Horn85}. Since the Frobenius norm is invariant under the Hermitian operation, we get $\|\mathbf {XX}^\mathsf H\|\leq \|\mathbf X\|^2$. Note that the approximations given in (\ref{eqn:DeltaOmegaTerm}) and (\ref{eqn:DeltaPhiTerm}) give an upper bound of the objective function of (\ref{eqn:optimization5}). By minimizing this upper bound, we also minimize the objective function.

Now, we replace the terms $\tr\left\{\mathbf G^\mathsf H\mathbf W\mathbf G\mathbf \Delta_\Omega\right\}$ and $\tr\left\{\mathbf F\mathbf F^\mathsf H \mathbf \Delta_\Phi\right\}$ in the robust transceiver problem (\ref{eqn:optimization6}) with the upper bounds defined in (\ref{eqn:DeltaOmegaTerm}) and (\ref{eqn:DeltaPhiTerm}) respectively. This is equivalent to setting
\begin{align}
\mathbf \Omega^\star=&\widehat{\mathbf \Omega}+\varepsilon_\Omega\mathbf I, \label{eqn:Omega_wc}\\
 \mathbf \Phi^\star=&\widehat{\mathbf \Phi}+\varepsilon_\Phi\mathbf I. \label{eqn:Phi_wc}
 \end{align}
 With these fixed worst-case matrices, the robust transceiver design problem reduces to
\begin{equation}
\begin{array}
[c]{ll}%
\underset{\mathbf F,\mathbf G}%
{\text{minimize }}& \underset{\|\mathbf \Delta_{H}\|\leq \varepsilon_H}{\text{max}} \tr\left\{\mathbf W\mathbf E\right\}\\
\text{subject to} & \tr\left\{%
\mathbf \Phi\mathbf F\mathbf F^\mathsf{H}\right\}  \leq P \\
& \mathbf \Omega=\widehat{\mathbf \Omega}+\varepsilon_\Omega\mathbf I, \quad \mathbf \Phi=\widehat{\mathbf \Phi}+\varepsilon_\Phi\mathbf I.
\end{array}
\label{eqn:optimization9}
\end{equation}

\textbf{Least Favorable Channel Error Matrix $\mathbf \Delta_H$:} Now, we find the worst-case channel estimation error $\mathbf \Delta_{H}$ following the maximization problem for given $\mathbf \Omega$ and $\mathbf \Phi$:
\begin{align}
\underset{\|\mathbf \Delta_{H}\|\leq \varepsilon_H}{\text{maximize}}  & \quad  \tr\left\{\mathbf A\mathbf \Delta_{H}\mathbf B\mathbf \Delta_{H}^\mathsf H\right\}+2\mathfrak{Re}\left\{\tr\left\{\mathbf C\mathbf \Delta_{H}\right\}\right\}.\label{eqn:optimization_Delta}
\end{align}
%
\par \textit{Lemma 2:} The least favorable channel estimation error for any given $\mathbf \Phi $ and $\mathbf \Omega$ has the following structure:
\begin{equation}
\mathbf \Delta_{H}^\star=\mathbf \Omega^{\frac{1}{2}}\widehat{\mathbf U}\widetilde{\mathbf \Delta}\widehat{\mathbf V}^\mathsf H\mathbf \Phi^{\frac{1}{2}}, \label{eqn:optimal_Delta}
\end{equation}
where $\widehat{\mathbf U} \in \mathbb C^{n_r \times n_r}$ and $\widehat{\mathbf V} \in \mathbb C^{n_r \times n_t}$ are defined in the SVD
\begin{equation}
\mathbf \Omega^{-\frac{1}{2}}\widehat{\mathbf H}\mathbf \Phi^{-\frac{1}{2}}=\widehat{\mathbf U} \left[\widehat{\mathbf \Gamma} \quad \mathbf 0_{n_r \times n_t-n_r}\right] \left[\widehat{\mathbf V} \quad \breve{\mathbf V}\right]^\mathsf H,\label{eqn:SVD2}
\end{equation}
and $\widetilde{\mathbf \Delta} \in \mathbb R^{n_r \times n_r}$ is a diagonal matrix with elements $\tilde{\delta}_{i} \geq 0$.
\begin{IEEEproof}
Due to limited space of this paper the detailed proof is included in \cite{Kaviani11c}. In order to explain our proposed transceiver optimization algorithm, here we only summarize the approach and results. The problem (\ref{eqn:optimization_Delta}) can be categorized as a \textit{trust-region subproblem} \cite{Stern95,Tao96}. The matrix-form restatement of this problem is given in \cite{Palomar10c}. It has been shown that the solution to this problem can be found by a minimization problem over an auxiliary variable $\vartheta\geq \lambda_{\max}(\mathbf A)\lambda_{\max}(\mathbf B)$ \cite{Stern95,Tao96}. The worst-case channel matrices coincides with the structure of the precoding and equalization matrices given in (\ref{eqn:F}) and (\ref{eqn:G}) using the worst-case interference plus noise and power shaping matrices defined in (\ref{eqn:Omega_wc}) and (\ref{eqn:Phi_wc}). As a result, $\tilde{\delta}_i$s are given by
\begin{align}
\tilde{\delta}_i=\frac{w_i\lambda_i\sigma_i(\gamma_i\lambda_i\sigma_i-1)}{\vartheta-w_i\lambda_i^2\sigma_i^2}, i=1,\ldots,n_r.
\label{eqn:delta}
\end{align}
Note that $\gamma_i, i=1,\ldots,n_r$ are the diagonal elements of $\widehat{\mathbf \Gamma}$ in (\ref{eqn:SVD2}). Recognizing $j=\text{argmax}_i\left(w_i\lambda_i^2\sigma_i^2\right)$, if $\vartheta> w_j\lambda_j^2\sigma_j^2$, then $\vartheta$ is the root of equation
\begin{equation}
\sum\limits_{i=1}^{n_r}\frac{w_i^2\lambda_i^2\sigma_i^2(\gamma_i\lambda_i\sigma_i-1)^2}
{\left(\vartheta-w_i\lambda_i^2\sigma_i^2\right)^2}=\widetilde{\varepsilon}_H^2, \label{eqn:error_power_condition}
\end{equation}
where $\widetilde{\varepsilon}_H=\frac{\varepsilon_H}{\|\mathbf \Omega^{\frac{1}{2}}\|\cdot \|\mathbf \Phi^{\frac{1}{2}}\|}$. If $\vartheta=w_j\lambda_j^2\sigma_j^2$, $\tilde{\delta}_j$ cannot be found from equation (\ref{eqn:delta}). 
Let
 \begin{equation}
\rho(\vartheta)=\sum\limits_{i\neq j}\frac{w_i^2\lambda_i^2\sigma_i^2(\gamma_i\lambda_i\sigma_i-1)^2}
{\left(\vartheta-w_i\lambda_i^2\sigma_i^2\right)^2}.
 \end{equation}
 Therefore, if $\rho(w_j\lambda_j^2\sigma_j^2)<\widetilde{\varepsilon}_H^2$, then $\tilde{\delta}_j=-\sqrt{\widetilde{\varepsilon}_H^2-\rho(\vartheta)}$. Otherwise, $\vartheta > w_j\lambda_j^2\sigma_j^2$ and it can be uniquely determined by (\ref{eqn:error_power_condition}).
\end{IEEEproof}

\subsection{Robust Transceiver Design}
Now, we can use the worst-case system matrices descriptions (\ref{eqn:Omega_wc}), (\ref{eqn:Phi_wc}), and (\ref{eqn:optimal_Delta}) and substitute into the robust transceiver design problem. Note that using the trust-region subproblems \cite{Stern95,Tao96} the resultant problem of finding worst-case channel estimation error $\mathbf \Delta_{H}$ becomes a minimization problem over an auxiliary variable $\vartheta$. The result can be compiled as follows:
\par \textit{Theorem 1:} The robust precoding and equalization matrices have the following structure:
\begin{align}
\mathbf F=&\left(\widehat{\mathbf \Phi}+\varepsilon_\Phi\mathbf I\right)^{-\frac{1}{2}}\widehat{\mathbf V}\mathbf \Sigma \label{eqn:B_optimal}
\\
\mathbf G=& \mathbf \Lambda \widehat{\mathbf U}^\mathsf H\left(\widehat{\mathbf \Omega}+\varepsilon_\Omega\mathbf I\right)^{-\frac{1}{2}}\label{eqn:A_optimal}
\end{align}
where
\par \textit{(i)} $\widehat{\mathbf U} \in \mathbb C^{n_r \times n_r}$ and $\widehat{\mathbf V} \in \mathbb C^{n_t \times n_r}$
are orthonormal matrices defined by the thin SVD \cite{Golub96}
\begin{align}
\left(\widehat{\mathbf \Omega}+\varepsilon_\Omega\mathbf I\right)^{-\frac{1}{2}}\widehat{\mathbf H}\left(\widehat{\mathbf \Phi}+\varepsilon_\Phi\mathbf I\right)^{-\frac{1}{2}}=&\widehat{\mathbf U}\widehat{\mathbf \Gamma}\widehat{\mathbf V}^\mathsf H \label{eqn:thin_SVD}
\end{align}
where $\widehat{\mathbf \Gamma} \in \mathbb C^{n_r \times n_r}$ is a diagonal matrix with diagonal elements of $\gamma_i \geq 0$,
\par \textit{(ii)} $\mathbf \Lambda$ and $\mathbf \Sigma$ are diagonal matrices of size $n_r$ with the diagonal elements of $\lambda_i, i=1,\ldots,n_r$ and $\sigma_i, i=1,\ldots,n_r$, respectively and they are obtained through solving the scalar optimization problem
\begin{equation}
\begin{array}
[c]{rl}
\underset{1\leq i \leq n_r}{\underset{\lambda_i,\sigma_i,\vartheta}{\text{minimize}}} & \sum\limits_{i=1}^{n_r}\frac{\vartheta w_i\left(\sigma_i\lambda_i\gamma_i-1\right)^2}{\vartheta-w_i\lambda_i^2\sigma_i^2}+
\sum\limits_{i=1}^{n_r}w_i\lambda_i^2+\vartheta\widetilde{\varepsilon}_H^2
\\
\text{subject to} & \vartheta \geq w_i\lambda_i^2\sigma_i^2, \quad i=1,\ldots,n_r
\\
& \sum\limits_{i=1}^{n_r}\sigma_i^2\leq P
\label{eqn:optimization10}
\end{array}
\end{equation}
\par \textit{(iii)} The optimum solutions for $\lambda_i$ and $\sigma_i$ can be obtained with respect to $\vartheta$ and a Lagrangian multiplier $\mu$ as follows
\begin{align}
\lambda_i=&\sqrt{X_i\sqrt{\frac{\mu}{w_i}}}
\label{eqn:lambda_i}\\
\sigma_i=&
\sqrt{X_i\sqrt{\frac{w_i}{\mu}}} 
\label{eqn:sigma_i}
\end{align}
where $X_i$ is a positive real root of a depressed quartic equation
\begin{align}
\varphi_i(X)=&\sqrt{\mu} w_i^2X^4-(2w_i\vartheta\sqrt{\mu}+w_i\sqrt{w_i}\gamma_i\vartheta)X^2 \nonumber\\
&(\gamma_i^2\vartheta+w_i)\sqrt{w_i}\vartheta X+\vartheta^2(\sqrt{\mu}-\gamma_i\sqrt{w_i})=0. \label{eqn:Quartic2}
\end{align}
If there is no real positive root, then $X_i=0$. The closed-form solutions for the roots of the quartic equation (\ref{eqn:Quartic2}) can be obtained using the Ferrari's method \cite{Spiegel08} and can be found in \cite{Kaviani11c}.
\begin{IEEEproof}
Due to limited space of this paper the detailed proof is included in \cite{Kaviani11c}. Here, we summarize briefly the approach used. We first substitute (\ref{eqn:Omega_wc}), (\ref{eqn:Phi_wc}), and (\ref{eqn:optimal_Delta}) into the original problem (\ref{eqn:optimization5}), and hence we simplify it to a minimization problem with respect to $\mathbf F, \mathbf G$ and the auxiliary variable $\vartheta$. Now, using Lemma 1, the optimal expressions for the precoding and equalization matrices are given by  (\ref{eqn:F}) and (\ref{eqn:G}) for any values of error matrices. Substituting expressions for $\mathbf F$ and $\mathbf G$ and the worst-case system matrices (\ref{eqn:Omega_wc}), (\ref{eqn:Phi_wc}), and (\ref{eqn:optimal_Delta}) into (\ref{eqn:optimization5}), we can convert the problem into a scalar optimization problem, which can be simplified to (\ref{eqn:optimization10}).
Notice that the maximization preserves the convexity, therefore this problem is a convex optimization problem with respect to $\mathbf G$ and $\mathbf F$ and consequently in $\lambda_i$ and $\sigma_i, i=1,\ldots,n_r$. By fixing $\vartheta$, we can solve the problem in $\lambda_i$ and $\sigma_i$. Next, the auxiliary variable $\vartheta$ will be updated following the Lemma 2. 
\end{IEEEproof}


\par Our closed-form solutions are functions of the auxiliary variables $\vartheta$ and $\mu$. Using dual decomposition concept from \cite{Palomar06a}, we can decompose the problem with outer loop optimization problems with respect to $\mu$ and $\vartheta$. These values can be updated using a subgradient algorithm \cite{Bertsekas03}. By differentiating the objective function in problem (\ref{eqn:optimization10}) with respect to $\vartheta$, we can obtain the subgradient direction for $\vartheta$ as
\begin{equation}
\Delta_\vartheta=\begin{cases}
\widetilde{\varepsilon}_H^2-\sum\limits_{i=1}^{n_r}\frac{w_i\lambda_i^2\sigma_i^2\left(\lambda_i\sigma_i\gamma_i-1\right)^2}%
{\left(\vartheta-w_i\lambda_i^2\sigma_i^2\right)^2} & \vartheta> w_j\lambda_j^2\sigma_j^2\\
\widetilde{\varepsilon}_H^2-\rho(\vartheta) & \vartheta=w_j\lambda_j^2\sigma_j^2
\end{cases}\label{eqn:v_subgradient}
\end{equation}
Similarly, by differentiation of the Lagrangian function of (\ref{eqn:optimization10}), we can get the subgradient direction for $\mu$ as
$\Delta_\mu=\sum_{i=1}^{n_r}\sigma_i^2-P.$ The robust transceiver optimization algorithm is summarized in Table~\ref{Tab:1}. The algorithm consists of two loops. The inner loop solves the convex scalar problem with respect to $\lambda_i$ and $\sigma_i$ and therefore it is convergent. The outer loop updates the auxiliary variable $\vartheta$ using a subgradient method, which is based on the strong duality of the trust region subproblem \cite{Stern95,Tao96} and consequently it is also convergent. The objective function of problem (\ref{eqn:optimization10}) is bounded and it is reduced in each iteration. Therefore, the algorithm in Table~\ref{Tab:1} is convergent.
\begin{table}
\centering
\caption{Robust Transceiver Optimization Algorithm}
\begin{tabular}{l}
\hline
Initialize $\sigma_i$s and $\lambda_i$s and $\mu>0$, $\vartheta >\max_i\left(w_i\lambda_i^2\sigma_i^2\right)$.\\
Perform thin SVD (\ref{eqn:thin_SVD}) to obtain $\gamma_i$s. \\
\textbf{Repeat} (subgradient loop of $\vartheta$)\\
\quad \quad Update $\vartheta\leftarrow \vartheta+\delta_\vartheta\Delta_\vartheta$ using (\ref{eqn:v_subgradient}).\\
\quad \textbf{Repeat} (subgradient loop of $\mu$)\\
\quad\quad  Form the quartic equation (\ref{eqn:Quartic2}) for $i=1,\ldots,n_r$. \\
\quad \quad Find its positive real root.\\
\quad\quad Find $\sigma_i$ and $\lambda_i$ using (\ref{eqn:lambda_i}) and (\ref{eqn:sigma_i}).\\
\quad\quad Update $\mu \leftarrow \mu+\delta_{\mu}\Delta_\mu$.\\
 \quad \textbf{Until} $\left|\sum_{i=1}^{n_r}\sigma_i^2-P\right|\leq \epsilon_0$\\
 \textbf{Until} satisfaction of (\ref{eqn:error_power_condition}) \\
 Replace $\lambda_i$s and $\sigma_i$s into (\ref{eqn:B_optimal}) and (\ref{eqn:A_optimal}) and find $\mathbf F$ and $\mathbf G$.\\
\hline
\end{tabular}\label{Tab:1}
\end{table}
\section{Numerical Results}
In this section, the performance of robust transceivers is evaluated numerically. The robust design guarantees a performance level for any point within the uncertainty region. Hence, the performance is displayed by the worst-case sum of MSE values, which are averaged over different system realizations. Each system realization is a result of a random generation of elements of the estimated system matrices (i.e. $\widehat{\mathbf H}, \widehat{\mathbf \Omega}^{\frac{1}{2}}, \widehat{\mathbf \Phi}^{\frac{1}{2}}$), which are i.i.d. Gaussian with zero mean. The uncertainty region is characterized by a parameter $0\leq \overline{\varepsilon} \leq 1$. In our simulations, it is assumed that $\sqrt{\overline{\varepsilon}}$ is the radius of the uncertainty region for each of the system matrices, when they are normalized by their Frobenius norms (i.e. $\varepsilon_H^2=\overline{\varepsilon}\|\widehat{\mathbf H}\|^2$, $\varepsilon_\Omega^2=\overline{\varepsilon}\|\widehat{\mathbf \Omega}\|^2$, and $\varepsilon_\Phi^2=\overline{\varepsilon}\|\widehat{\mathbf \Phi}\|^2$ ). The non-robust transceivers consider the estimated system matrices as the actual system matrices and are discussed in Lemma 1 and \cite{Kaviani11b}. The worst-case  estimation error matrices have been given in Section \ref{Sec:ImperfectCSI}. Note that only a solution of a special case of our problem is available in the literature, which includes uncertainty of the channel matrices $\mathbf H$ only. For this special case, our algorithm performs as the algorithm in
\cite{Wang11} while it is less complex (by optimizing the precoder and equalizer jointly and reducing the problem to a quartic equation). It is shown for the special case (uncertainty of $\mathbf H$) that our algorithm performs as well as SDP methods with much lower complexity using an iterative approach.
\begin{figure}
\centering
\includegraphics[width=3.7in,height=2.5in]{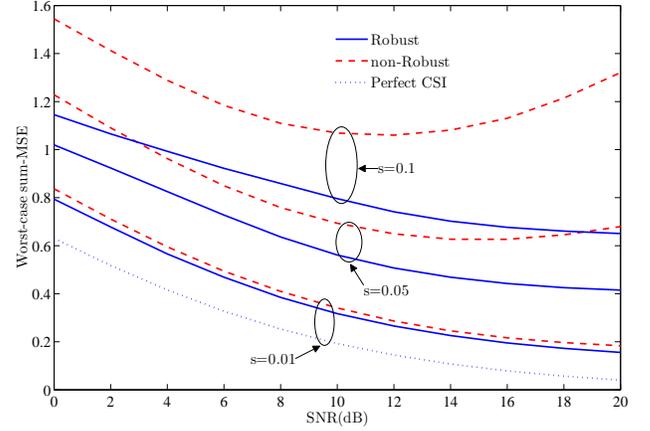}
\caption{Comparison of the proposed robust design, the non-robust design \cite{Kaviani11c}, and the transceiver design when system matrices are perfectly known (perfect CSI) for $n_t=n_r=2$. \label{Fig:1}
}
\end{figure}
\par Fig.~\ref{Fig:1} shows the comparison of robust and non-robust designs for different values of $\overline{\varepsilon}$, i.e. the size of uncertainty regions. The sum-MSE of the transceivers obtained with perfect knowledge of system matrices is also given as a baseline.
\begin{figure}
\centering
\includegraphics[width=3.7in,height=2.5in]{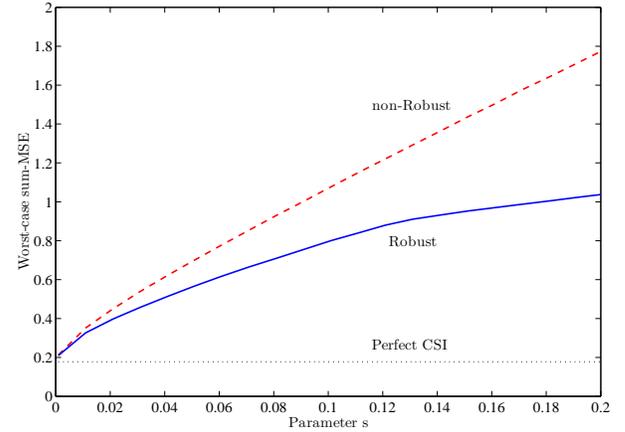}
\caption{Comparison of different transceiver designs with respect to the size of uncertainty region $s$ for $n_t=n_r=2$. \label{Fig:2}
}
\end{figure}
Fig.~\ref{Fig:2} explicitly illustrates the performance of the robust and non-robust design with respect to the size of the uncertainty region. As expected, the performance of the robust transceivers degrades at a much lower pace with increase of signal-to-noise-ratio (SNR) and the size of uncertainty region $\overline{\varepsilon}$ compared to the non-robust transceivers.

\section{Conclusions}
We have designed the robust transceivers when the channel matrix, interference plus noise covariance matrix, and power shaping matrix (system matrices) are \textit{all} imperfectly known to the transmitter. The closed-form expressions for the precoder and equalizer have been found. This involves finding the worst-case system matrices first and then simplifying the problem to a scalar convex form. The solution to this optimization problem can be described in a form of a depressed quartic equation, the closed-form expressions for roots of which are known. Finally, we have proposed an iterative algorithm to obtain robust transceivers, which is significantly less complex compared to SDP-based alternating optimizations. Moreover, accounting for imperfect knowledge of all system matrices enables for the extension of our approach to the multiuser scenario, which is the subject of our current work \cite{Kaviani11c}.
\bibliographystyle{IEEEtran}
\bibliography{IEEEabrv,MyBib}

\end{document}